\documentclass[fleqn,10pt]{SelfArx} % Document font size and equations flushed left
\usepackage[toc]{appendix}
\usepackage{lipsum} % Required to insert dummy text. To be removed otherwise

\definecolor{color1}{RGB}{0,0,90} % Color of the article title and sections
\definecolor{color2}{RGB}{0,20,20} % Color of the boxes behind the abstract and headings

\usepackage{hyperref} % Required for hyperlinks
\hypersetup{hidelinks,colorlinks,breaklinks=true,urlcolor=color2,citecolor=color1,linkcolor=color1,bookmarksopen=false,pdftitle={Title},pdfauthor={Author}}

\usepackage{booktabs}
\usepackage{makecell}
\usepackage{tabularray}
\usepackage{array}
\usepackage{xcolor}
\UseTblrLibrary{booktabs}
\usepackage{tabularx}
\usepackage{caption}
\newcolumntype{L}{>{\raggedright\arraybackslash}X}

\JournalInfo{Draft Version} % Journal information
\Archive{} % Additional notes (e.g. copyright, DOI, review/research article)

\PaperTitle{Framework of Voting Prediction of Parliament Members} % Article title

\Authors{Zahi Mizrahi\textsuperscript{1}, Shai Berkovitz\textsuperscript{1}, Nimrod Talmon\textsuperscript{2}, Michael Fire\textsuperscript{1}}
\affiliation{\textsuperscript{1}\textit{Department of Information Systems Engineering, Ben-Gurion University, Beer Sheva, Israel}} % Author affiliation
\affiliation{\textsuperscript{2}\textit{Department of Industrial Engineering and Management, Ben-Gurion University, Beer Sheva, Israel}} % Author affiliation

 % Keywords - if you don't want any simply remove all the text between the curly brackets
\newcommand{\keywordname}{Keywords} % Defines the keywords heading name

\Abstract{Keeping track of how lawmakers vote is essential for government transparency. While many parliamentary voting records are available online, they are often difficult to interpret, making it challenging to understand legislative behavior across parliaments and predict voting outcomes. Accurate prediction of votes has several potential benefits, from simplifying parliamentary work by filtering out bills with a low chance of passing to refining proposed legislation to increase its likelihood of approval. In this study, we leverage advanced machine learning and data analysis techniques to develop a comprehensive framework for predicting parliamentary voting outcomes across multiple legislatures. We introduce the Voting Prediction Framework (VPF) - a data-driven framework designed to forecast parliamentary voting outcomes at the individual legislator level and for entire bills.
VPF consists of three key components: (1) Data Collection - gathering parliamentary voting records from multiple countries using APIs, web crawlers, and structured databases; (2) Parsing and Feature Integration - processing and enriching the data with meaningful features, such as legislator seniority, and content-based characteristics of a given bill; and (3) Prediction Models - using machine learning to forecast how each parliament member will vote and whether a bill is likely to pass. The framework will be open source, enabling anyone to use or modify the framework. 
To evaluate VPF, we analyzed over 5 million voting records from five countries - Canada, Israel, Tunisia, the
United Kingdom and the USA. Our results show that VPF achieves up to 85\% precision in predicting individual votes and up to 84\% accuracy in predicting overall bill outcomes. These findings highlight VPF's potential as a valuable tool for political analysis, policy research, and enhancing public access to legislative decision-making.}

\begin{document}

\flushbottom % Makes all text pages the same height

\maketitle % Print the title and abstract box

%%\tableofcontents % Print the contents section

\thispagestyle{empty} % Removes page numbering from the first page

%----------------------------------------------------------------------------------------
%	ARTICLE CONTENTS
%----------------------------------------------------------------------------------------

\begin{figure*}[hbtp]
   \includegraphics[width=\linewidth]{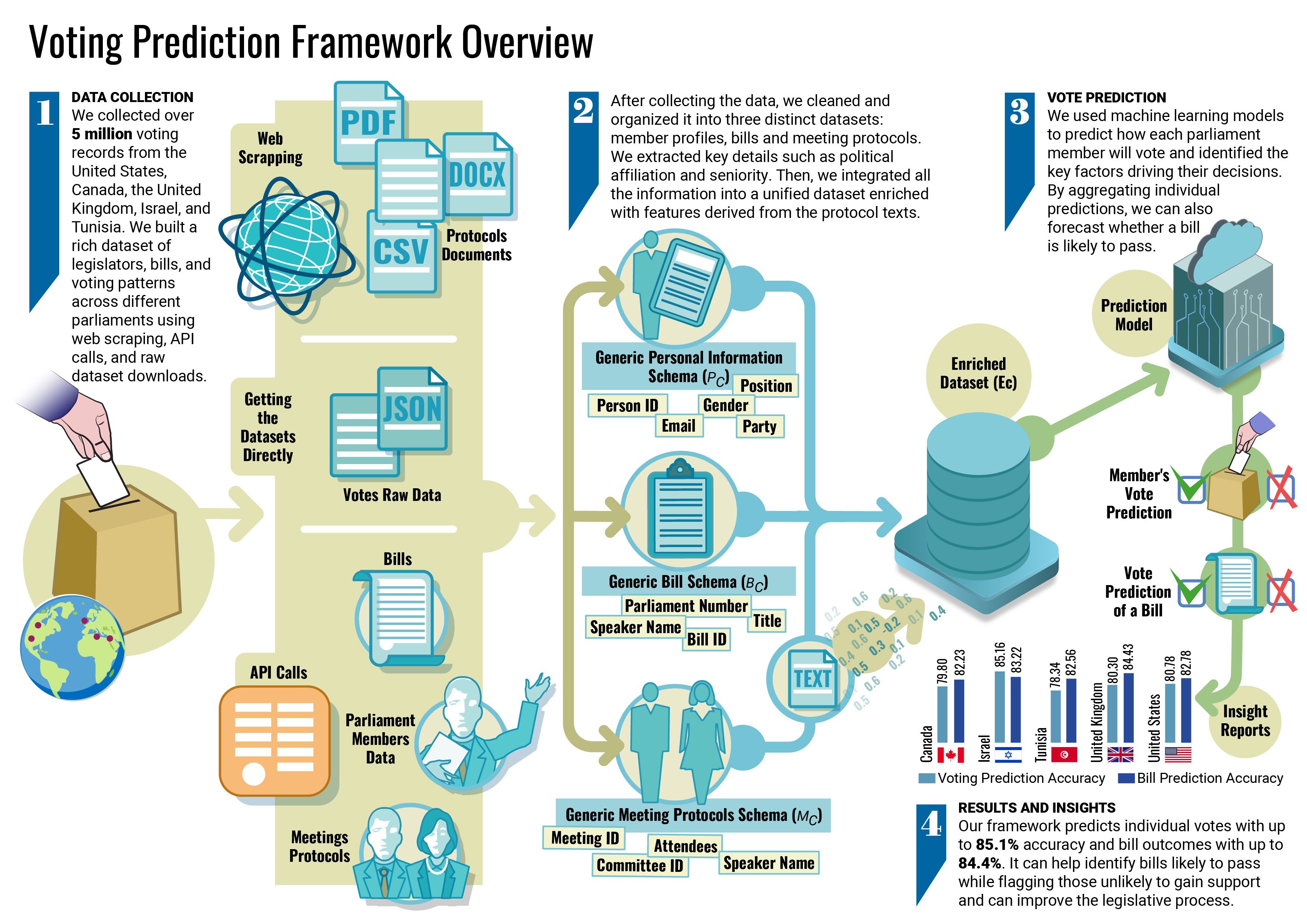}
   \caption{Framework Overview}
   \label{fig:framework_flow}
   \caption*{Voting Prediction Framework (VPF) overview. VPF contains three main components: The first is data collection of parliamentary records, second is parsing and feature calculation and integration, and the last - vote prediction using machine learning models. }
\end{figure*}

% -----------------------------------------------------

\section{Introduction} % The \section*{} command stops section numbering

Political scientists and researchers have analyzed the outcome of congressional roll-call votes for decades. Patterns in congressional voting behavior, reflected in Open Government Data (OGD), have proven valuable in predicting the results of these votes ~\cite{gerrish2011predicting}. The analysis of roll-call data helps gain deeper information about legislators (such as political leanings and ideologies) and predict future legislative decisions ~\cite{clinton2012using}. 

Open Government Data (OGD) typically refers to public records and information sourced from governmental entities or activities, encompassing important sectors such as transportation, infrastructure, education, and health~\cite{hendler2012us}. 
These datasets are made available for reuse and redistribution, primarily free of charge or at a nominal cost~\cite{geiger2012open}. It is widely believed that OGD can be helpful for social and economic progress~\cite{krishnamurthy2016liberating, berends2017analytical, huyer2020economic}. For example, during the initial stage of the COVID-19 pandemic, researchers used government data to understand the social and economic impact of the pandemic on civil society~\cite{gonzalez2021open}.

Various projects and systems have been trying to process and analyze governmental data using open government information over the years. One type of project focuses on analyzing public datasets, such as the GovTrack.us project, which collects the voting records of the USA Congress and analyzes statistical insights on the roll-call votes~\cite{act2015govtrack}. Another type of project focuses on collecting and processing governments' data by public APIs or web scraping and making the data more accessible. For example, Canada and the US parliaments' OGD portals~\cite{quarati2019open} contain over 
25,000\footnote{Data UK: {\url{https://data.gov.uk}}} and 287,000\footnote{Data Gov: 
 {\url{https://open.canada.ca}}} datasets, respectively.

In this study, we present the Voting Prediction Framework (VPF), our generic framework that offers data analysis on voting records and prediction of the parliament votes on an offered bill. Using this analysis, we can identify bills that have a high chance of getting accepted in the voting process, as well as some bills that should not even be considered due to their low expected chance of passing. Additionally, our framework can help to investigate the unexpected voting behavior of parliament members.

VPF contains three main components (see Figure~\ref{fig:framework_flow}): The first component collects relevant parliamentary data from different countries using various methods, including using APIs or building web crawlers. 
The second component is parsing and feature integration. The framework processes and parses the collected data, enriches it to complete missing values, and extracts significant dataset features. Those features include the political affiliation of the parliament member, the importance rank that indicates the level of seniority of the parliament member's position in the parliament, and the number of references on a given subject in past bills and protocols. In addition, we add an embedding layer of the bill content as features to the framework, representing the integration of the bill content into the parliament member data. 
The last component uses machine learning models to predict how each parliament member would vote on a bill proposal and explain which factors contributed the most to the result of the model.

To test and evaluate our framework, we collected data from the parliaments of five countries: Canada, Israel, Tunisia, the United Kingdom and the USA (see Table ~\ref{table:votes_data_sum}). The data contains personal information on parliament members (such as their position in the parliament or their political affiliation), their sayings in meeting protocols, and their voting records. We collected thousands of personal information records and over 5 million voting records in 3 different languages (English, Hebrew, and Arabic). Then, we extracted features from this dataset and used a prediction model to predict the voting outcome. Subsequently, we performed an large-scale experiment to evaluate the framework's accuracy in predicting the legislators' voting The results show high precision of 79\%, 85\%, 78\%, 80\%, 80\% in Canada, Israel, Tunisia, the United Kingdom and the United States, respectively (see Figure ~\ref{fig:bar_charts}). Also, we calculated the bill prediction by grouping all the votes for each bill and analyzing if the prediction of the majority of the votes is aligned with the test result. The bill prediction shows accuracy of 82\%, 83\%, 82\%, 84\%, and 82\% in Canada, Israel, Tunisia, the United Kingdom, and the United States, respectively  (see Table ~\ref{table:bill_prediction_results}). Furthermore, by investigating false negative results, we found examples of anomalous voting behavior of some individual parliamentarians and some critical issues that conflict with their government (for example, the carbon tax conflict in Canada ~\cite{harrison2012tale} ). Moreover, by using SHAP values that evaluate the importance rank of each feature in the dataset, we can analyze what influences the voting result. 

The primary contribution of this research is developing a platform capable of analyzing rich parliamentary data (for example, voting records) and predicting future voting based on parliamentary activity and individual information regarding members. We can divide the contribution as follows:

\begin{itemize}[noitemsep]

 \item \textbf {Prediction of Likelihood of Legislative Bills Passing}: The VPF framework can help prioritize bills and estimate the likelihood of passing a bill. It can also reduce the time and effort consumed by parliament, making it a robust tool for decision-making and resource allocation in legislative processes.

 \item \textbf{Generic Platform: } Each country has its parliament, each of which has its unique way of publishing voting records and bills. Our framework is generic and not limited to a particular country. In addition, the framework can support additional features tailored to specific countries.

 \item \textbf{Open Framework}: The source code of the framework will be available upon publication, allowing anyone to enhance, modify, or inspect the implementation of the framework. It helps with knowledge sharing and enables global comparison of different parliaments.  

 \item \textbf {Detecting Voting Behavior}: Our framework helps to detect abnormal voting behavior patterns of parliamentary members by investigating false negative voting results. Those insights may sometimes be used to inform the public about the real intentions of their parliamentarians on critical issues.
 
 \end{itemize}

The rest of this article is structured as follows. Section  ~\ref{sec:related_work} provides an overview of studies and methods for data mining on legislative voting. 
Section ~\ref{sec:methods} describes our methodology for building the framework and the experiments we conducted to evaluate our method in Section ~\ref{sec:experiments}. Section ~\ref{sec:results} details our obtained results. In Section ~\ref{sec:discussion}, we discuss the obtained results and what we can infer from them. Lastly, Section ~\ref{sec:conclusion} presents the paper's findings and outlines future research directions. 

%------------------------------------------------
\section{Related Work}
\label{sec:related_work}
In our research, we use open parliament data  to predict legislative voting.  In this section, we will show the significance of open government data (see Section ~\ref{subsec:open_parliament_data}),  provide an overview of related works to existing open parliament framework (see Section ~\ref{subsec:open_framework}), detail the legislative process and its complication (see Section ~\ref{subsec:legislative_process}) and reviews different approaches to legislative voting prediction (see Section ~\ref{subsec:data_mining_voting}).

\subsection{Open Parliament Data}
\label{subsec:open_parliament_data}
Open Government Data (OGD) plays a crucial role in improving the government's communication with its citizens. Additionally, it ensures that the valuable information held by governmental authorities is being used correctly to boost social and economic benefits~\cite{fetisova2020role}.
In recent years, there has been a rise in the number of open data movements, with the
two primary aims of transparency and data reuse. These movements have allowed multiple
platforms, such as 'data.gov.uk', and 'data.gov', to emerge. These platforms have
led the way for easily accessible government information and activities, enabling citizens
and stakeholders to obtain valuable information ~\cite{quarati2019open}.

Initially, combating corruption was among the primary goals of establishing open government data (OGD) ~\cite{attard2015systematic}. 
 Corruption is a global issue that seriously harms the economy and society, affecting people’s lives and often infringing fundamental human rights. The
democracy of many countries worldwide is undermined by deep-rooted corruption ~\cite{stephenson2015corruption},
which also affects financial growth. While the total economic costs of corruption cannot
be easily calculated, the 2014 European Commission Anti-Corruption Report states that
corruption can be estimated to cost the European Union economy 120 billion euros per year 
~\cite{hoxhaj2019eu}. 

The Western world's constantly evolving open government policy has been actively seeking emerging technologies to enhance communication channels between governmental systems and the public. Additionally, this policy aims to optimize the usage of valuable
social and economic information within governmental authorities. At both the federal and regional tiers, public sector information technologies are being strategically used  to enhance the operational efficiency of administrative processes within these governmental bodies ~\cite{hulstijn2017open}.

There are three primary motivations for the founding of most government data initiatives.
According to Ubaldi et al.~\cite{ubaldi2013open}, unlocking data in the digital age promotes transparency, access to information, and participatory governance.
Additionally, it is essential that governments seek
feedback from the public on the usefulness, relevance, and accessibility of their data to enable continuous improvement. Furthermore, numerous government officials recognize the
need to observe the real-time performance and impact of public services and policies concerning citizens. They can construct relevant data or
use available data to improve the service experience,  provide tools, and provide incentives. An example of this can be seen through opportunities to
participate in professional roles on online social networks, allowing individuals to offer advice and information to the public. Therefore, governments need to recognize the value of their audience's sources of real-time data and information sharing and engage relevant stakeholders outside public organizations while using them to create value ~\cite{ubaldi2013open}.

Moreover, several federal data monitoring and conservation projects have been conducted, which are scientifically hosted on government databases in the US and other
countries ~\cite{carrara2015creating}. Today, there is a trend shift among governments to try to be more “accessible" ~\cite{ubaldi2013open}. However, it is known that simply providing open government data does not automatically contribute significant value to society ~\cite{janssen2012benefits}. The literature often cites the numerous potential benefits of OGD ~\cite{ubaldi2016rebooting, carrara2015creating}.  However, these benefits are still believed to materialize only if the data is used. Thus, a concrete understanding of barriers that prevent the use of OGD to produce public value is essential. Therefore, a framework is needed to guide the use of OGD efficiently and effectively~\cite{carrara2015creating}.

\subsection{Open Parliament Frameworks}
\label{subsec:open_framework}
In recent years, there has been a greater understanding of OGD, and a growing number of bodies, public institutions, and governments have acknowledged and adapted to this change.  Today, many OGD-related datasets are available to the general public across various countries, offering increasingly relevant and practical information ~\cite{von2016benchmarks}.

For example, there is the OGD portal project ~\cite{quarati2019open}, publishing portals that provide visible data on the political section of each country. For example, the UK\footnote{Data UK: {\url{https://data.gov.uk}}} and the US\footnote{Data Gov:{\url{https://data.gov}}} parliaments’ portals contain over 25,000   and 287,000 datasets, respectively.

Today, many different open code initiatives aim to utilize available governmental datasets. The \textit{Openkamer} code project\footnote{Openkamer:\url{https://github.com/openkamer/openkamer}} provides insight into the Dutch parliament by extracting parliamentary data from several external sources. It visualizes the data in a web application, including legislative proposals, queries, political parties, gifts to parliament, and more. Similarly, in the UK,\footnote{Public Whip Project: \url{https://www.publicwhip.org.uk}} an independent non-governmental project,  web-scrapes House of Commons and House of Lords debate transcripts to enable the public to monitor and influence voting patterns.

Additionally, similar projects have been implemented in France and Canada. 
In France, the \textit{Senapy} Python project\footnote{Senapy: https://github.com/regardscitoyens/senapy} scrapes data from the French Senate website.\footnote{Senate Website: \url{https://senat.fr}} Meanwhile, in Canada, the \textit{coalition analyzer}\footnote{\url{https://github.com/oliversno/coalition-analyzer}} involves a code that submits open API requests for the Canadian House of Representatives to calculate correlations in the votes of different parties. 

In Israel, the \textit{Open Knesset project}\footnote{\url{https://oknesset.org}} mines all the activities of the Israeli parliament (Knesset) from the official Knesset website to enable tracking of voting, legislation, and committee activities.
In 2020, an open-source website was developed to inspect and explore the Israeli parliament.\footnote{\url{https://github.com/SgtTepper/BetaKnessetWeb}} The tool allows users to search for various topics using keywords, find parliament members’ votes, and view the activities of all parliament members, including the schedule of their committees. Using this tool enables citizens to explore all the prominent topics discussed during parliament sessions and examine politicians’ interests and agendas. 

In the USA, Voteview\footnote{\url{https://voteview.com/}} allows users to view every congressional roll-call vote in American history on a map of the USA and on a liberal-conservative ideological map, including information about the ideological positions of voting Senators and Representatives. Moreover, GovTrack.us\footnote{\url{https://www.govtrack.us/}} also helps to monitor the USA Congress by collecting statistical analyses on the roll-call votes~\cite{act2015govtrack}. Lastly, another tool is LocalView~\cite{barari2023localview}, an open-source data portal containing the most extensive existing dataset of local government meetings in real-time. It has been used to study local policy-making in the USA, which has been expensive to study at scale. 

In 2023, Berkovitz et al.~\cite{berkovitz2022open} presented a framework for analyzing parliaments' data. This framework is generic and enables the analysis of large-scale public governmental data of parliament protocols from any country, to find anomalous meetings and detect dates of events that affect the parliament's functionality. This study used the framework to collect over 64,000 parliament protocols from over 90 committees from three countries. \textit{Our study aims to extend the collection capabilities and add additional capabilities of analyzing parliamentary data}, primarily regarding legislative bills data accessibility and prediction methods on the data.

\subsection{The Legislative Process}
\label{subsec:legislative_process}
\begin{figure}[h]
    \centering
    \includegraphics[width=\linewidth]
    {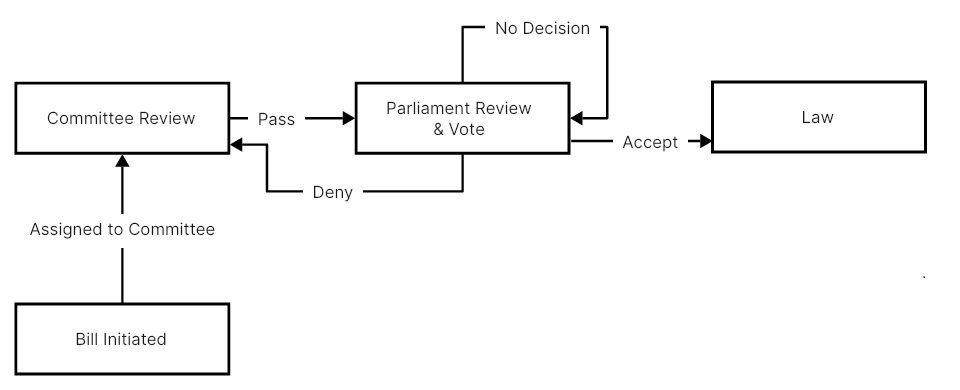}
    \caption{Illustration of the legislative process}
    \label{fig:legislative_process}
\end{figure}

The foundation of many democratic countries is formed by parliaments and state legislatures that represent the interests of their constituents. Adding, changing, and removing laws is the primary responsibility of Parliament and the way of making changes in their state. The legislative process is referred to as the method by which a legislative proposal is turned into an act.
The legislative processes vary in different countries, but the major principles are similar ~\cite{karvonen2007legislation}. In the USA,  the process starts with one or more legislators identifying the need for a new law or an amendment ~\cite{sinclair2016unorthodox}. The idea of a bill can also originate from a constituent public official or an interest group. Following the legal requirement, the relevant legislatures write a declaration text for the proposed legislation, known as a bill. This bill is then assigned to a committee for study and review. Committees are made up of a subset of parliament members. Other relevant parliament members are informed about this bill so they can make any amendments or alterations.
For a bill to become law, it goes through a reviewing process where it can "die" on different pages. If the bill gets a passing vote, it is sent to Parliament with a date for its vote, and the same process repeats. In some countries, like the USA, the bill must be passed by a simple majority before moving on to the Senate. It must pass through several readings or get different approvals in other cases. If the bill gets approval, it is then made into law. However, if the bill is not approved, it can be returned to the committee for further consideration. The process is described in Figure~\ref{fig:legislative_process}.

As previously mentioned, the legislative process is complicated, and a bill's approval is influenced by mainly two sets of effective factors~\cite{karimi2019multi}. The first set is ideological factors, which are known to have an essential role, for example, in the USA Congress~\cite{kraft2016embedding} and come from both the parliament members as well as the ideology of the bills, whose values and beliefs are represented in the content of the bills. The second set of influential factors are social factors related to the coalition affiliation of representatives and how their past voting behavior intertwines with the other representatives. It is known that parliament members are more likely to follow their political affiliation when voting ~\cite{andeweg2011pathways} and that there is a connection between the bills and the voting behavior of the parliament members on similar bills in the past ~\cite{butikofer2015strategic}.

\subsection{Data Mining on Legislative Voting}
\label{subsec:data_mining_voting}
Several studies used data mining on parliamentary data. For example, some used Natural Language Processing (NLP) methods to detect political ideology, sentiment, position-taking, perspectives, and the level of agreement and disagreement between politicians.
For example, in 2016, Rheault et al.~\cite{rheault2016measuring} used NLP methods to calculate the emotional polarity in debates of the British Parliament, using parliamentary data spanning a century in the United Kingdom parliament. Abercrombie et al. ~\cite{abercrombie2020sentiment} published a literature review that included 61 studies focusing on the automatic analysis of sentiment, expressed opinions, and positions taken by speakers in parliamentary (and other legislative) debates on various topics.

Over the years, some studies have used data mining methods on legislative voting since voting patterns in Parliament offer one of the few sources of behavioral data to study members of parliaments. According to Hug et al.~\cite{hug2013parliamentary}, for optimizing the information stemming from parliamentary voting, the full context of the parliament members' choices has to be considered since votes can be influenced by various factors like their party, their background, etc.

In 2016, Bräuninger et al.~\cite{bateman2016ideal} developed a roll-call voting model that considers legislators' policy and non-policy incentives. This model analyzed thousands of roll-call votes from German state legislatures and the British House of Commons. This study showed that tactical incentives may be more important than policy incentives.
Predicting the behavior of legislators, which represents their voting behavior, largely relies on roll-call data, i.e., historical voting records, to estimate the political preference of legislators. Some of the studies used additional data, such as the social activity of parliament members~\cite{mou2021align, vafa2020text}. Gerrish and Blei~\cite{gerrish2011predicting} used topics associated with USA Congress Bills and then predicted the likelihood of a Parliament member voting in support of a bill.

In 2019, Godbout and Hoyland~\cite{godbout2011legislative} analyzed legislative votes from the Canadian Parliament, based on the relation between governing and opposition parties, to find voting behavior. Eidelman et al. ~\cite{eidelman2018predictable} built a prediction model for roll-call votes in the USA Congress, predicting whether a bill would pass the preliminary stage to a full-body vote. They used contextual information about the legislators and legislatures with the bill text. However, that research does not consider whether the bill would become law.  

In 2022, Ankit Pat et al.~\cite{pal2022deepparliament} introduced a benchmark dataset that gathered bill documents and metadata from 1986 to 2022 and performed various bill status classification tasks, including voting result prediction. The study used different deep learning models ranging from RNN to BERT. 

There are some studies on voting predictions that are not necessarily related to parliamentary voting. For example, Persing et al. ~\cite{persing2014vote} built a system for voting prediction on comments in social polls. Some other examples are prediction of presidential elections using data from Twitter ~\cite{liu2021can, rita2023social} or surveys representing voters' opinions~\cite{murr2021vote}. In 2023, Kavallos et al. ~\cite{kavallos2023parliament} used the Greek Parliament's content of debates, speeches, and discussion to classify the parliamentary processing to the owning political party. 

Recently, parliamentary alignment with political parties has been examined through analyses of open government data. This research often focuses on estimating defection voting rates, which occur when members of parliament vote against the majority of their party or in contrast to their party's agendas. In 2024, Mai et al.~\cite{mai2024loyal} conducted a regression analysis on dissenting voting behavior using data from the German Bundestag. The study results showed behavioral differences related to pre-parliamentary socialization and political engagement with the party.

In our study, we used similar data types for prediction analysis, but with a different innovative approach to voting prediction compared to the research studies above.

%--------------------------------------------------
\section{Methods}
\label{sec:methods}
The primary purpose of our study is to implement a framework for the prediction of votes of parliament members by analyzing and processing different data sources, including protocols, documents, and parliament information, which was obtained using open government data APIs and raw protocol data from web scraping.

The following subsections provide an overview of the development of the generic framework for collecting structured information from parliamentary bodies. Our method consists of the following steps (see Figure~\ref{fig:framework_flow}): First, we collect the parliamentary data and the voting records from publicly available sources. Then, we preprocess the records into structured datasets (see Section~\ref{subsec:data_collection_and_parsing}. Next, we utilize the data, adding significant features based on the data, and by using machine learning algorithms, we construct classifiers that can predict the voting of a given bill (see Section~\ref{subsec:data_analysis}) and evaluate them. Then, we use the prediction results and calculate the prediction accuracy for each bill to see if our model can be generalized in predicting the outcome of legislative bills.   Lastly, we evaluate the prediction results, analyze the votes the framework failed to predict, and calculate how much each feature contributed to the model's prediction. In addition, we analyze the negative results from the framework and give reasonable explanations or discover weaknesses of the framework (see Section~\ref{subsec:validation}).

\subsection{Data Collection and Preprocessing}
\label{subsec:data_collection_and_parsing}

Data collection is the first primary phase of our voting framework. Given a country with parliament data that is available online, we build a web crawler that collects all the information into datasets. For many parliaments, typically, there are three different data collections (parliament members' information, bills, and votes) and other sources of meeting protocol documents. To collect data, we can use various collection methods, such as official APIs like ODATA API\footnote{\url{https://www.odata.org/}} and web scraping applications, which extract relevant data according to the setting for each data source. 

It is necessary to convert the collected unstructured data into structured data to create the datasets. For each country, we develop a dedicated parser to construct each data collection. Since the data can be collected in various formats across different countries, the parser is designed to support multiple input formats and transform the data into a predefined schema. We utilize tools like regular expressions to extract the values according to the defined schema datasets, especially in cases where the input contains free text instead of straightforward values. The data collected and extracted includes personal information about parliament members, details about legislative bills, voting records, and the content of meeting protocols. 

In order to deal with missing values in the parsed data during pre-processing. We implemented several solutions to infer missing information from other fields or alternate sources to address missing values. For example, if a parliament member's full name is missing in the votes dataset, it could be deduced from the personal information dataset (see the unified schema in Table \ref{tab:votes_schema}). Another method for handling missing values includes using data from the preceding and following rows, in cases where missing data is incremental or collected in ascending order (like time or ID). In cases where these methods cannot resolve missing values, we manually update the missing values.

 To create the voting dataset, we construct each voting session in the voting records to contain several attributes, including the bill being voted on, the voting date, and the voting preference of each representative. The voting data are similar to roll-call votes, with four different types of vote results: (1) Yes, indicating parliament members' approval of the bill; (2) No, indicating parliament members' disapproval of the bill; (3) Abstention, indicating parliament member's decision not to participate in the voting; and (4) Obstruction, similar to abstention, but with the difference that it counts for quorum effects, i.e., the minimum number of voting members required to be present for the vote to be valid, while obstruction does not account for it. In the collected data of votes, not all countries have all these types of vote results, but have at least yes or no vote results.

\begin{table*}[hbtp]
    \centering
    \caption{Summary of the voting data records collected from 5 countries.}
    \begin{tblr}{|c|c|c|} 
        \toprule
        Country & Number of Years & Number of Voting Records \\
        \midrule
        \textbf{Canada} & 15 & 683,942\\
        \textbf{Israel} & 23 & 816,756\\
        \textbf{Tunisia} & 2 & 368,262 \\
        \textbf{United Kingdom} & 8 & 273,948\\
        \textbf{United States} & 19 & 2,850,296\\
        \bottomrule
  \end{tblr}
   \label{table:votes_data_sum}
\end{table*}

We define a generic schema for each type of dataset. This schema, also called the unified schema, contains all the attributes that provide a single, consistent view of an entity and helps organize data from multiple sources. Also, this schema can be extended with fields necessary only for the specific parliament to support the unique properties of different parliaments. 

Namely, by analyzing the collected unstructured data of a given country $C$, we construct the following datasets (see Appendix ~\ref{sec:tables_schema}):%%\footnote{If required, the unified schema can be extended for to contain additional information.}:
\begin{enumerate}
    \item \textbf{The Personal Information Dataset} contains various personal details about parliament members, denoted as $P_C$.
    \item \textbf{The Bills Dataset} contains information on all the bills (including subject, the member who proposed it, etc.), denoted as $B_C$. 
    \item \textbf{The Raw Votes Data Dataset} contains information on the vote of each parliament member on a given vote, denoted as $V_C$. 
    \item \textbf{The Meeting Protocols} details discussion of the parliament on various topics, denoted as $M_C$.
\end{enumerate}
%Information about the schema of each dataset can be found in Appendix ~\ref{sec:tables_schema}.  

\begin{table*}[hbtp]
    \centering
    \caption{Generic Data Votes Schema ($V_X$)}
    \begin{tblr}{|c|c|} 
        \toprule
        Features & Description \\
        \midrule
        Country             &  Country name  \\
        Vote ID             & Unique identifier of the vote \\ 
        Parliament Number   & The number of the parliament occurred \\ 
        Session ID          & List Unique identifier of the voting session \\
        Vote Date           & The date of the vote \\
        Total For           & Number of members who voted for the proposal\\
        Total Against       & Number of members who voted against the proposal\\
        Member ID           & ID of the Parliament member\\
        Member Name         & Parliament member full name\\
        Party ID            & Unique identifier of the member's party\\
        Party Name          &  Name of member's party \\
        Member Gender       &  The Gender of the member \\
        Is Current          &  A boolean indicating if the member is in the parliament now\\ 
        Vote Result         &  Kind of vote (1-for, 2-against, 3-abstain, and 4-did not vote) \\
        \bottomrule
  \end{tblr}
  \label{tab:votes_schema}
\end{table*}

\subsection{Data Analysis}
\label{subsec:data_analysis}
After the preprocessing phase, we compiled a table containing all the relevant information to predict the votes for each country. 

Initially, for a given country defined as $C$, we take the raw data vote dataset $V_C$ with the required information from all the bills of the parliaments $B_C$. This allows us to access all the comprehensive details about the subject and the proposal for each vote within the parliament. Then, we enrich the data for each bill proposal and vote using the information from the other datasets, creating the enriched dataset $E_C$. Subsequently, we added some significant features to the dataset $E_C$: 

\begin{itemize}[noitemsep]
  \item \textbf{Political Affiliation:} A boolean value that indicates if the parliament member belongs to a party belonging to the coalition (or on the same side). The way political affiliation is structured within a parliament is influenced by various factors. In coalition governments, political affiliation is often formed based on ideological proximity, meaning parties on similar parts of the spectrum are likelier to vote together. The feature calculates whether the current party of the parliament member is in alliance with the coalition or not. 

    \item \textbf{Importance Rank:} Calculating numeric rank based on the position of the parliament member in the parliament. We believe influential positions within the parliament may influence other members' votes. The rank $R$ ranges from $R_H$ (signifying a vital position, such as prime minister) to $R_L$ (signifying a standard parliament member without additional responsibilities).\footnote{The importance rank range is selected as the longest hierarchy in the parliament.} The rank is determined based on a predefined dictionary that assigns a level of importance to each existing parliamentary position. This information is then calculated against the position of each parliament member. 

     \item \textbf {Opinion on Subject:} It is essential to understand the opinion of each parliament member on the voting subject to predict their vote. To achieve this, we used the data from the parsed protocols. We searched for any references to the bill's subject within the content of the protocols in which the parliament member participated. Then, we count the number of references to the bill's subject in the parsed protocol. This value is added as a feature to the dataset.

    \item \textbf{Bill Embedding:} We create an embedding of the bill protocol description using pre-trained embedding models. Embeddings transform textual data into numerical representations, where similar texts are positioned closer together in a high-dimensional vector space. These pre-trained models are commonly used for tasks like clustering~\cite{xie2016unsupervised}, semantic search~\cite{muennighoff2022sgpt}, or text classification~\cite{moreo2021word}. They allow the mapping of sentences and paragraphs into dense vector representations that capture semantic meaning. By using these embeddings, our framework can better understand the context of bill proposal descriptions. For example, similar bill proposals will have embeddings that are located next to each other in the vector space.\footnote {It is worth noting that there are various methods and models for generating embeddings, and the choice of embeddings depends on the language of the data, as embeddings are language-dependent.}
    
\end{itemize}

Next, using the extraction of relevant attributes, the complete dataset was prepared as input for a multi-class classification model (the number of classes is the number of voting result types, as explained in Section ~\ref{subsec:data_collection_and_parsing}). The features used in this predictive model include a wide range of contextual information, such as the parliament member's position, political affiliation, individual opinion on key issues, and similarity of bill proposals (the schema can be found in Table~\ref {tab:enriched_schema}). This approach may be used for each parliament since this information is not fine-tuned to a specific parliament, and the assumption made on the features is generally valid for parliaments worldwide.

We used various machine learning models and algorithms on the features dataset, including Decision Trees ~\cite{hoens2012building}, Random Forest ~\cite{prinzie2008random}, Multi-layer Perceptron (MLP) Classifier ~\cite{popescu2009multilayer}, SVM~\cite{joachims1998making}, Gaussian Naïve Bayes ~\cite{jahromi2017non}, and XGBoost ~\cite{chen2016xgboost}. We run each machine learning model on the dataset, used Accuracy, F1-score \cite{sokolova2006beyond} to measure the accuracy of each prediction model, and then choose the model with the highest accuracy using AUC.\footnote{We used all the models with their default configuration, except for XGBoost models, which were used with the parameters early stopping $rounds=30$, $n estimators=1000.$}. Then, we selected the model with the highest accuracy to evaluate bill prediction accuracy. This was accomplished by aggregating all votes for each bill and assessing whether the predicted results matched the results for most votes. The result is the number of bill proposals for which the model successfully predicts the majority of the votes, divided by the total number of bill proposals. 

\subsection{Model Evaluation}
\label{subsec:validation}
%In order to check the quality of our framework, validation is important.
We splitted the data into the training and testing datasets to evaluate our prediction model. We utilized the training dataset to train the model and the testing dataset to evaluate the model.

Random splitting is a commonly used method that randomly divides the dataset into training and test sets \cite{reitermanova2010data}. However, in our case, random splitting may lead to bias and does not accurately test the ability to predict the voting on new bills, because it is important to capture the past chronological order of events on a given subject of a bill before predicting future bill proposals. 
Therefore, we used time-series splitting to validate our framework. This type of data splitting was used in past studies for different tasks, such as rumor detection ~\cite{mu2023s} or anomaly detection in URLs~\cite{harikrishnan2019time}.

Because some data does not have date or time attributes with fine-grained resolution, it is challenging to keep splitting as 75\% training and 25\% testing. We used the date resolution that gives us the closest resolution of 75\%-25\% to training and testing parts (see Table ~\ref{table:votes_dataset}).
After splitting the data into training and testing datasets, we measured the accuracy and F1-score of the different models for every country (see Table ~\ref{table:results_table}). For each prediction, we calculated the accuracy and the F1-Score and drew a ROC (receiver operating characteristic) curve of the model with the highest accuracy for each country. The ROC curves helped us understand how well the models balance true positives (correct predictions) and false positives (wrong predictions). The closer a curve is to the top-left corner, the better the model makes correct predictions. The AUC (Area Under the Curve) value tells us how good the model is overall, where a higher AUC means better performance. We used the AUC values to compare the results between the countries. 
Lastly, we used SHAP ~\cite{vstrumbelj2014explaining} to explain the output of the machine learning models. SHAP (SHapley Additive exPlanations) values are used to explain the output of a model by attributing each feature’s contribution to the final prediction. 
SHAP values offer a clear, feature-by-feature breakdown of a model's predictions, making complex algorithms more transparent and interpretable.
We utilized the SHAP values to measure the contribution of each feature to the result and explain which features were the most important for the prediction.

%Also, for each bill proposal, we calculated the probability of bill prediction rather than predicting each individual vote. This was done by grouping all the votes for each bill and determining whether the predicted outcomes aligned with the actual results for the majority of the votes.

Additionally, we examine false negative examples in the test set to gain insight into
the anomalies, which can be made from two main causes. 
The first is unexpected behavior by parliament members, indicating unknown factors influencing their votes, such as coalition agreements or extraneous considerations. 
The second cause could be errors in the prediction model, resulting from low-quality data, missing information, or issues within the framework.

%--------------------------------------------------

\section{Experiments}
\label{sec:experiments}
We conducted the framework’s experiments on data from five countries: Canada, Israel, Tunisia, the United Kingdom, and the United States. Data collection varied across countries in terms of the methods used, the time range of the records, and the number and types of data sources (see Table ~\ref{table:votes_data_sum})--additionally, the set of features we used as the prediction model input differs across countries.  In the following subsections, we will detail the data collection, parsing, and feature selection processes for each country.

\subsection{Canada}
\label{subsec:canada}
In Canada (CA), we collected committee data from the House of Commons, the lower house of the Parliament of Canada. We collected data by scraping the official website of the House of Commons.\footnote{\url{https://www.ourcommons.ca/en}} There were no exported APIs to obtain the committee data, so we implemented a designated scraper for this website. We also implemented a scraper and parser for the voting records data, taken also from this website. We collected over 680,000 voting records on legislative bills from 2008 to 2023. This data contains information on five parliaments, 795 different parliament members, and 13 provinces.

In Canada's voting records, there were only two types of voting decisions - yes or no. Abstention votes are dismissed from the collected data. Because the data collected did not contain the meeting protocol data, we only used the importance rank, the political affiliation advanced features, and all the voting records schema. For the time-series split, we used data before 2023 as the training data, and the rest as the test set, which divided the data into 81\%-19\% training-test datasets.

\subsection{Israel}
\label{subsec:israel}
We collected personal information about parliament members from the official Knesset website.\footnote{\url{https://main.knesset.gov.il/}} It was mainly accomplished by utilizing web scraping scripts that extracted the necessary information based on site-specific settings, and through the official ODATA API for information about Parliament members. For document of meeting protocols, we scraped the metadata for each document from the protocol search page of the Knesset. Then, we downloaded each document using this metadata and created a schema for information from the document protocols. We collected 13,500 documents representing the last 25 years of parliamentary proceedings. For voting records, we scraped the metadata for each vote from the voting search page of the Knesset and then iterated for each parliament and downloaded the results. We collected 800,000 of legislative votes, representing the last 25 years of parliamentary proceedings. This dataset includes more than 10,000 voting sessions of legislative bills deliberated in the Knesset, from 1999 to 2022. 

 We parsed all the relevant data and the documents to structured text using open-source libraries.\footnote{We used the python-docx library, which is utilized to handle Microsoft Word documents.} We also parsed metadata about the documents (connection to committee type, date, author, etc.). Then, we analyzed the opinion of parliament members on the subject using those documents as a feature of the model. For bill embedding, since the data is in Hebrew, we used AlephBert~\cite{seker2021alephbert}. 
 
 In Israel's voting records data, there were only three types of voting decisions - Yes, No, or Abstention. As for the time-series split data, we used all the data before 2018 April 1st as the training data, and the rest as the test set, which divides the data into 75\%-25\% training-test datasets.

\subsection{Tunisia}
\label{subsec:tunis}
 In Tunisia, we collected parliamentary data from the Marjad Mahles website,\footnote{\url{https://majles.marsad.tn/}} which shows the parliamentary data of Tunisia. This data can be collected in Arabic or French only. Also, there is no recent history or activity of the parliament data, only from 2012 to 2014. As a result, we collected over 368,262 voting records in Arabic from 2012 to 2014, representing over 1700 roll-call votes. 

In Tunisia's voting records, there were four types of voting decisions - Yes, No, Abstention, or Obstruction. For the time-series split, we used the data before May 2014  as the training dataset, and the rest as the test dataset, which divides the data into 78\%-22\% training-test datasets.

\subsection{United Kingdom}
\label{subsec:united_kingdom}
In the UK, we collected data on parliament members, legislative bills, and voting records from the UK parliament's committee API.\footnote{\url{https://committees-api.parliament.uk/}} UK Parliament is the official website of the United Kingdom Parliament, which has two houses that work on behalf of UK citizens to make and shape effective laws and decisions. However, there is not enough historical data exported from their API. We collected over 850,000 voting records from 2006 to 2023, covering five parliaments and including data on 923 parliament members.

We collected and parsed the data from the 'parliament.uk' API into a structured table with a schema.  We used  political affiliation and bill embedding features using the entire MiniLM-L6-v2 version of `sentence-transformers`.

In the United Kingdom voting records, there were only three types of voting decisions - Yes, No, or Abstention. For the time-series split, we used the data before August 2026 as the training dataset, and the rest as the test dataset, dividing the data into 77\%-23\% training-test datasets.

\subsection{United States}
\label{subsec:united_states}
In the USA, we collected 
data from the committees of the US House of Representatives by scraping the GovInfo website.\footnote{\url{https://www.govinfo.gov/}}  The GovInfo also provides free public access to official publications from all three branches of the Federal Government. GovInfo includes access to many documents, including information on publications or collections of content, or a view of a list of collections. The documents were collected from January 1999 to May 2021. Also, we used ProPublica API\footnote{\url{https://www.propublica.org/}} to get personal information about members of the parliament, the bills, and the voting records of their congress. We collected more than 2 million voting records from the congress, from 2005 to 2024, including information from 10 parliament numbers and 1126 parliament members from 50 provinces. 

We collected the data from the ProPublica API and parsed it into a structured table with a schema. We used the feature of political affiliation. Additionally, we used the bill embed feature using the entire MiniLM-L6-v2 version of `sentence-transformers`.

In the USA Voting records, there were only three types of voting decisions - Yes, No, or Abstention. For the time-series split, we used all the data before  2023 as the training dataset, and the rest as the test dataset, which divides the data into 73\%-27\% training-test datasets.

%--------------------------------------------------

\section{Results}
\label{sec:results}
In the experiment, we collected data from 5 different countries. The data contained more than 2.8 million protocols over 19 years from the US parliament across 21 committees; nearly 700,000 voting records over 15 years from the Canadian parliament across 44 committees; over 800,000 voting records over 23 years from the Israeli parliament; nearly 280,000 voting records over 8 years from the United Kingdom Parliament; and a collection of 368,000 voting records over 2 years from Tunisia Parliament (see Table~\ref{table:votes_data_sum}).

\begin{table*}[hbtp]
    \centering
    \caption{Summary of dataset, split by training part and testing part.}
    \begin{tblr}{|c|c|c|c|} 
        \toprule
        Country & Total Records in Dataset & Training Size & Testing Size \\
        \midrule
        \textbf{Canada} & 683,492 & 554,772 (81\%) & 128,720 (19\%)\\
        \textbf{Israel} & 816,756 & 612,567 (75\%) & 204,189 (25\%)\\
         \textbf{United Kingdom} & 867,523 & 669,346 (77\%) & 198,226 (22\%)\\
         \textbf{Tunisia} & 368,262 & 289,347 (78\%) & 78,915 (22\%) \\
        \textbf{United States} & 1,048,600 & 741,020 (73\%) & 287,580 (27\%)\\
        \bottomrule
  \end{tblr}
  \label{table:votes_dataset}
\end{table*}

Table ~\ref{table:results_table} shows the performance of 6 machine learning models - XGBoost, Decision Tree, Random Forest, MLP Classifier, and SVM on voting prediction across five countries. Overall, XGBoost performs the best, consistently achieving the highest accuracy and F1 scores, followed by Random Forest models. Israel shows the highest accuracy - 85.166\%.

\begin{figure*}[hbtp]
    \centering
    \includegraphics[width=\textwidth]{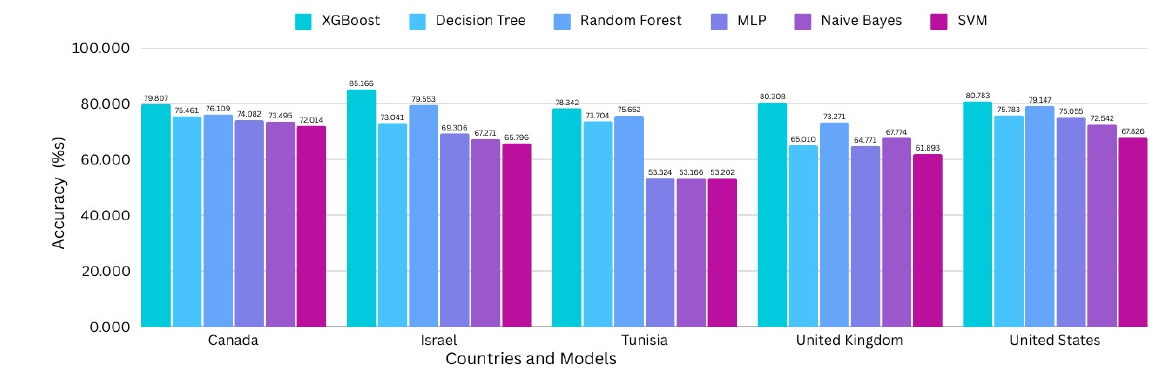}
    %%\Description{Bar chart of the accuracy of the framework by each model in Canada, Israel, Tunisia, United Kingdom and United States}
   \caption{Experiments Result Summary of Vote Prediction}
   \caption*{Bar chart describing the accuracy of each model (by percents), separated by countries. We can see that XGBoost classifier is the model that shows the highest accuracy in all countries. }
   \label{fig:bar_charts}
\end{figure*}

\begin{figure*}[hbtp]
    \centering
    \includegraphics[scale=0.75]{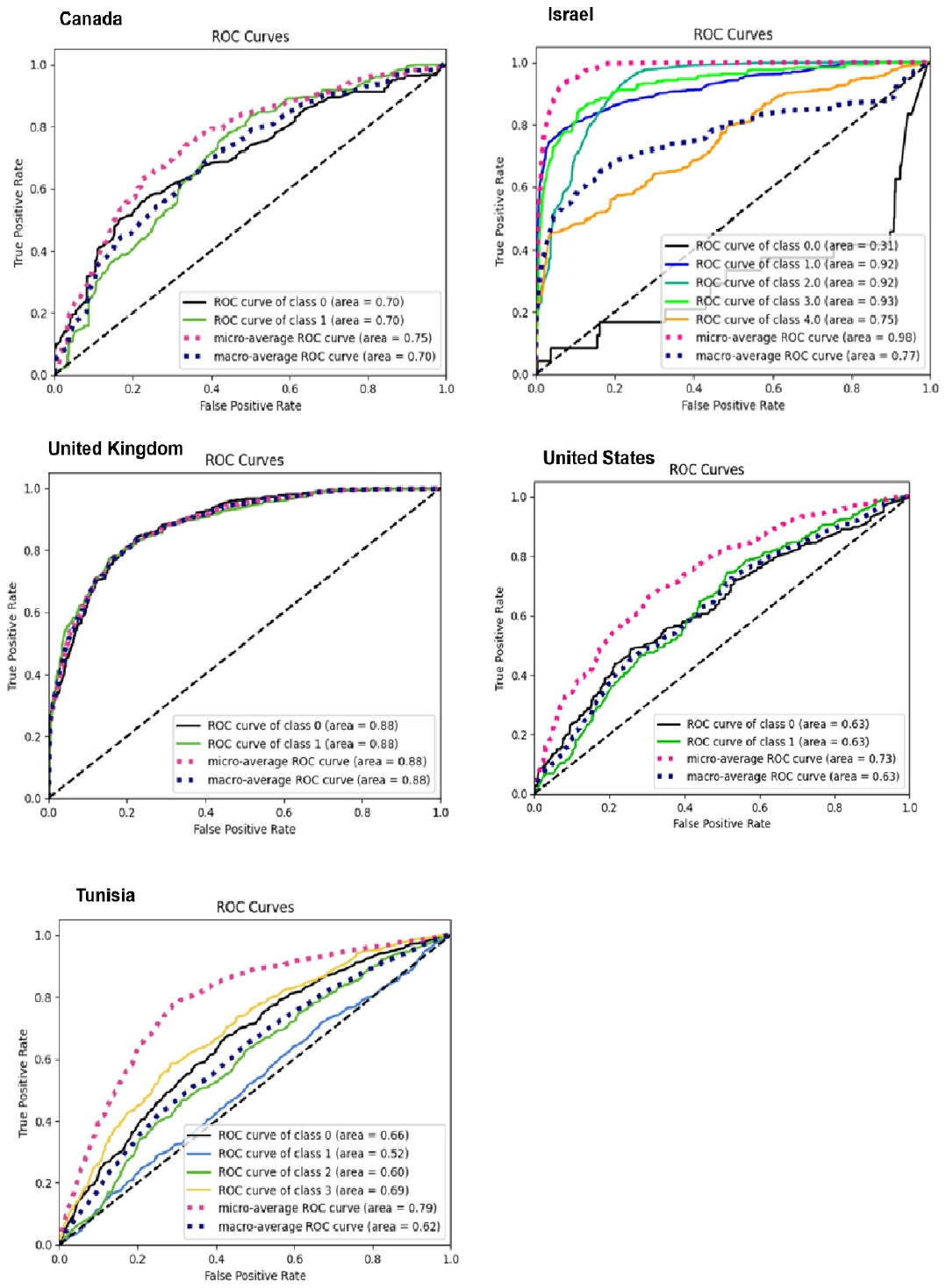}
    %\Description{ROC Curves of Canada, Israel, Tunisia, United Kingdom and United States}
   \caption{ROC Curves}
   \caption*{ROC Curves of Canada, Israel, United Kingdom, United States and Tunisia (where class 0 = Yes; class 0 = No; class 2 = Abstention; class 3 = Obstruction; class 4 = Did not Vote). We can see that the micro-average AUC (that does take imbalance into account) are 0.75, 0,98, 0.88, 0.73, 0.79 respectively.} 
   \label{fig:roc_curves}
\end{figure*}

\begin{figure*}[hbtp]
    \centering
    \includegraphics[width=\linewidth]{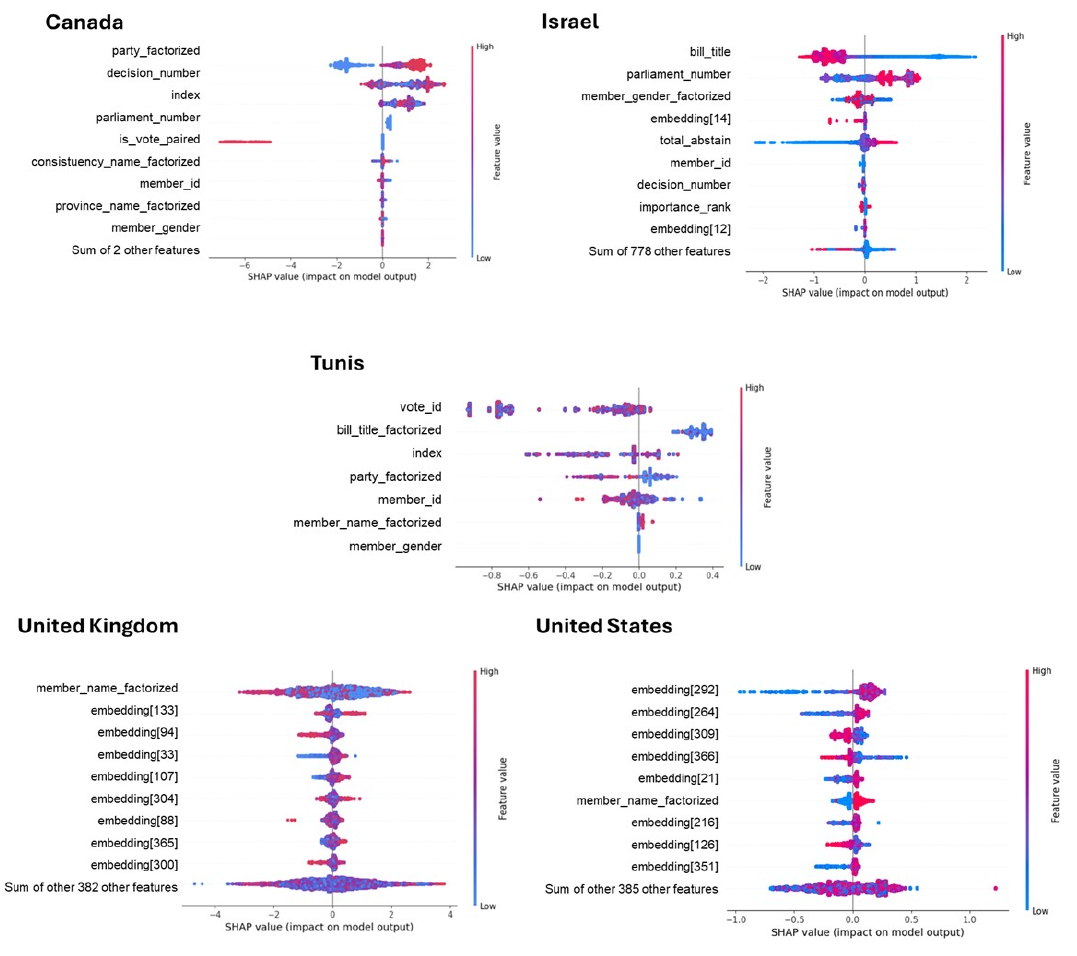}
    %%\caption&{SHAP Beeswarm visualization of Canada, Israel, Tunisia, United Kingdom, and United States}
    \caption{SHAP values in Beeswarm visualization of Canada, Israel, Tunisia, United Kingdom, and United States. }
   \label{fig:shap_graphs}
\end{figure*}

The results are summarized in Figure ~\ref{fig:bar_charts}, and the ROC Curves graphs, along with the AUC scores, are presented in Figure ~\ref{fig:roc_curves}. Additionally, the SHAP values are visualized in Figure ~\ref{fig:shap_graphs}.

 In addition, we calculated the probability of bill prediction at the level of each bill proposal, rather than for individual votes. This was done by aggregating all votes associated with each bill and determining whether the predicted outcomes matched the actual results for the majority of votes. The results are presented in Table~\ref{table:bill_prediction_results}, with a split of outcomes for passed and rejected bills shown in Table~\ref{table:bill_prediction_results_by_passed}. The accuracy rates of bill prediction by country are as follows: 82\% in Canada, 83\% in Israel, 82\% in Tunisia, 84\% in the United Kingdom, and 82\% in the United States.

\begin{table*}[hbtp]
    \centering
    \caption{Experiment Results of Bill Prediction.}
    \begin{tblr}{|c|c|c|c|} 
        \toprule
        Country & Total Number of Bills & Successful Predicted Number of Bills & Accuracy \\
        \midrule
        \textbf{Canada} & 411 & 338 & 82.238\% \\
        \textbf{Israel} & 2713 & 2258 & 83.228\% \\
        \textbf{Tunisia} & 367 & 303 & 82.561\% \\
         \textbf{United Kingdom} & 437 & 369 & 84.439\% \\
        \textbf{United States} & 302 &  250 & 82.781\%\\
        \bottomrule
  \end{tblr}
  \label{table:bill_prediction_results}
\end{table*}

In our analysis of the negative results from the framework, we identified several notable anomalies.

In Israel, the model had difficulty predicting votes for the "Yesh Atid" party during the 23rd Knesset when they were part of the "Kahol Lavan" coalition, as well as for "Hareshima Hameshutefet." There were also inconsistencies within other homogeneous parties. For instance, Yehuda Glick (Likud) and Mordechai Yogev (Habayit Hayehudi) voted against a bill commemorating the Kfar Qasim massacre, despite the overall support from their coalition. Similarly, on a legislative bill from 2017 regarding the operation of public transportation on Saturday (Shabbat), the model incorrectly predicted that 3 Meretz Party Members Raz Musi, Tamar Zandberg and Ilan Gilon to vote against it, when in reality, they supported it. Notably, the model successfully predicted the stance of other Meretz members in this case.

In the United States, the model struggled to differentiate between Republican votes for and against legislation, especially when members deviated from party lines. A significant anomaly arose with the "Pandemic is Over Act," presented on January 30, 2023, to terminate the COVID-19 public health emergency. The model failed to predict the Republican resistance to this law, particularly regarding the motion-to-recommit (MTR). According to House rules, an MTR provides one final opportunity for the House to debate and amend a rule before the Speaker orders a final vote on passage. In reality, the majority of MTRs are quickly dismissed by the party in power, making them a tool often used by the opposition to send a political message before a final vote.

In Canada, prediction errors were widespread concerning tax-related bills, often misjudging the agendas of the Conservative and Liberal parties. For example, during the 44th Parliament on January 31, 2024, a proposed bill to exempt the agriculture sector from a federal carbon tax was incorrectly predicted to face Conservative opposition and receive Liberal support. Those limitations of the framework mirror similar mistakes identified in other tax-related votes. Additionally, the model struggled to account for rare exceptions within party voting patterns, such as Philip Lawrence and Chris Warkentin breaking from Conservative opposition in favor of funding for museums and economic development during the 44th Parliament on December 7, 2023.

%--------------------------------------------------

\section{Discussion}
\label{sec:discussion}
Based on the results presented (see Section ~\ref{sec:results}), we can observe the following: 

First, a large amount of parliamentary data is accessible online. We used different methods described in Section ~\ref{subsec:data_collection_and_parsing} to collect more than 5 million voting records from five countries, from average of 11.4 years (see Table~\ref{table:votes_data_sum}), along with more than 70,000 government meeting protocols and parliament member information from the past 20 years. This dataset can help us study government activities and trends over periods. It is worth mentioning that the data collected was incomplete; in some of the records we had missing values. Also, there are some missing records in the dataset. Although the missing data, our framework succeeds in predicting the votes of parliament and has proven to be versatile for voting records. While we have shown that the framework is functional across various parliaments, it may also be worth noting that manual adjustments are required  to gather data from different parliamentary systems and to optimize its use in the data analysis phase. These adjustments may involve fine-tuning feature calculations or model parameters. Nevertheless, we have designed the voting prediction framework to be adaptable and generic, allowing its extension to accommodate other parliamentary contexts.
Additionally, despite the inherent differences in context and schema among the data collected from different countries, it was feasible to standardize the data into a unified schema and context. The unified schema enhances the usability of the proposed framework for comparative analysis of parliamentary actions and voting behavior across countries.

Secondly, we showed notable enhancements to the dataset for predicting voting patterns and bill passage chances. We compared different machine learning models and explained the results. We highlighted the predictive power of a limited set of features in predicting voting behavior. Despite the modest quantity of features, our analysis demonstrated a voting prediction accuracy ranging from 79\% to 85\% for parliament members on voting (see Table ~\ref{table:results_table}) and from bill prediction accuracy ranging from 82\% to 85\% (see Table ~\ref{table:bill_prediction_results}).

Thirdly, we analyzed which features influence the model’s prediction the most using SHAP values for each country. From the SHAP values, we can see that different countries have different key factors affecting voting behavior. 
In Canada, some of the most important factors are party affiliation and decision number. This result suggests that party loyalty and procedural aspects of voting are crucial in predicting votes in Canada and that individual characteristics play a smaller role, as was concluded in Godbout et. al~\cite{godbout2011legislative}.

In Israel, we can see that the key features are the bill title, the parliament number, and the member's gender alongside embeddings (representing past voting behavior). This  result indicates that the nature of the bill and the parliamentary context influence the voting in Israel. 
In the United Kingdom, some of the most important features are member name followed by text embedding features, which means that voting in the United Kingdom is highly influenced by individual voting history rather than general party trends. 
In the United States, similar to the United Kingdom, member names and embeddings are important, meaning past individual voting history is a strong predictor of voting. 
In Tunisia, we see lower SHAP values overall, indicating that voting behavior is more uncertain and less structured than in other countries. 

Fourthly, we can analyze using different machine-learning models for voting prediction. The results indicate that XGBoost performs better than the other models. Other models, like SVM and Naïve Bayes, generally have the lowest accuracy, while Decision Trees and Random Forest perform well but outperform XGBoost. Our results show that Israel has the highest AUC values (AUC of 0.98), followed by the United Kingdom (AUC of 0.88), suggesting that the XGBoost model is particularly well suited for use in these countries.

Fifthly, from the discovery of examples of anomalies in the prediction outcomes, several key conclusions can be drawn regarding the challenges of our VPF framework. First, our framework performs well when party members vote consistently along their party lines but struggle with individual deviations. This suggests that while party ideology is a strong predictor, individual parliament members may vote based on personal beliefs, regional interests, or strategy, making predicting votes more difficult. Secondly, the framework may oversimplify ideological alignments, assuming uniform stances on specific issues (for example, the tax-related bills anomaly in Canada). This observation highlights the need for deeper contextual analysis beyond broad party affiliation.

It is important to acknowledge several limitations of the framework and methods outlined in this study. During the research, we made certain assumptions about the data we used - the first is that the datasets consumed from government platforms are accurate. The second is that in cases where we have never observed past parliamentary actions of a member, our framework might predict voting with less certainty. Also, our predictions were based on the fact that the parliaments use majority voting. However, the framework can extend to different voting rules. 

Despite the analytical value of voting prediction patterns, they present significant opportunities for transparency and raise ethical and social concerns. One notable risk is the potential misuse by adversarial entities, such as lobbyists, political consultants, or interest groups, who might exploit predictive insights to influence legislators or manipulate voting results, undermining democratic processes. More broadly, there exists a risk that legislative proposed legislation bills could be strategically crafted to align with model prediction, prioritizing passage likelihood over policy goals or social warfare. These concerns highlight the need for robust ethical guidelines and transparency standards to mitigate the risks of strategic abuse and maintain the integrity of the legislative process. 

Additionally, our framework can analyze fundamental attributes like parliament member names or bill titles and calculate features such as political affiliation or rank of importance. However, it does not include specific features for different parliaments, which could reveal important details about the voting process and the unique traits of each parliament. We believe that adding these features could improve the accuracy of the prediction. Also, the SHAP values analysis shows which significant features in each parliament could lead to better results. For example, we discovered that past voting history is a dominant factor in the United States and the United Kingdom, implying their parliaments are more stable and voting behaviors are based on historical trends and features.
Another example is that Tunisia and Israel's voting patterns are influenced by bill content, meaning issue-based content features may influence voting. Also,  the personal data we collected about parliament members comes only from the parliamentary dataset. It does not include their agendas or personal beliefs. 

Lastly, our framework can be used successfully to pinpoint legislation bills that stand a substantial chance of approval in the voting process while also highlighting certain bill proposals that should be dismissed entirely. 

%--------------------------------------------------
\section{Conclusion and Future Work}
\label{sec:conclusion}
The study presents an innovative and flexible framework for voting predictions, designed to collect parliamentary data from various sources, including meeting protocols, voting records, and more. This framework collects, parses the data, and analyzes it through calculated features applied to classification machine learning models. To enhance the credibility and accuracy of our predictions, we gathered a large-scale dataset of over 5 million voting records from five distinct countries. We derived insightful features by parsing these records and their associated protocols and utilized prediction models to predict parliament members' votes on specific bills. Finally, we conducted a comparative analysis of different machine learning models to identify the one with the highest accuracy. This comprehensive analysis allows us to highlight bills with a promising chance of acceptance in the voting process, while also recognizing those that may not even be taken into consideration.

This study has numerous future research directions. Firstly, we can expand the external sources the framework uses. An example of future research is leveraging social media network information to evaluate the opinion of the citizens aligned with specific political parties and integrating this information as a feature for vote prediction. Secondly, we can broaden the framework's scope by incorporating additional countries such as Australia or France, and show comparative analyses of similar bills across different legislative bodies. 
Thirdly, we suggest integrating other methods from the NLP tools, such as sentiment analysis tools~\cite{medhat2014sentiment} or BERT topic modeling~\cite{grootendorst2022bertopic} to extract further insights into the topics and content of each bill. Supporting these types of additional features would enable more comprehensive comparisons among parliamentary bodies. Moreover, as the volume and complexity of data increase, there is a potential for further research into the world of prediction using deep learning models. Lastly, a website can be developed to visualize real-time prediction analysis of votes using our framework.

The VPF framework can collect worldwide open parliamentary data for government research on voting patterns and trends. It can also be utilized to offer valuable assistance in prioritizing legislative bills. This prioritization aids in optimizing parliamentary efforts and resources in the legislative process. Furthermore, the framework's ability to estimate the likelihood of bill passage can improve parliamentary operation efficiency.

%--------------------------------------------------
\phantomsection

\section*{Acknowledgments}
Thanks to Valfredo Macedo Veiga Junior (Valf) for designing the infographic illustration.

The authors used language enhancement tools, including ChatGPT and Grammarly, to improve the clarity and readability of this research paper. %----------------------------------------------------------------------------------------
%	REFERENCE LIST
%----------------------------------------------------------------------------------------
\phantomsection
\bibliographystyle{unsrt}
\bibliography{sample}

%----------------------------------------------------------------------------------------
%	APPENDIX LIST
%----------------------------------------------------------------------------------------

\appendix
\section*{Appendix: Dataset Schema \& Results}
\label{sec:tables_schema}

* Represents an optional feature;

\setcounter{table}{0}
\renewcommand{\thetable}{S\arabic{table}}

\begin{table*}[hbtp]
    \centering
    \caption{Generic Personal Information Schema ($P_C$).}
    \begin{tblr}{|c|c|} 
        \toprule
        Features & Description \\
        \midrule
        Person ID           & Unique identifier of the person \\
        Parliament Number   & The number of parliament of the data \\
        First Name          & First Name of the parliament member \\
        Last Name           & Last Name of the parliament member \\
        Gender              & Gender of the parliament member (Male or Female) \\
        Email               & E-Mail of the parliament member \\
        Party               & Party of the parliament member \\
        Position            & Position (Job) of the parliament member \\
        \bottomrule
  \end{tblr}
  \label{tab:personal_information_schema}
\end{table*}

\begin{table*}[hbtp]
    \centering
    \caption{Generic Bill Schema ($B_C$).}
    \begin{tblr}{|c|c|} 
        \toprule
        Features & Description \\
        \midrule
        Bill ID             & Unique identifier of the person \\
        Parliament Number   & The number of parliament of the data \\
        Title               & The title of the bill \\
        Committee ID*       & The ID of the Committee which the bill was proposed \\
        Date                & The time which the bill was proposed \\
        Description*        & Description of the bill (the content) \\
        Speaker Name        & The ID of the Parliament Member that proposed the bill \\
        \bottomrule
  \end{tblr}
   \label{tab:bill_schema}
\end{table*}

\begin{table*}[hbtp]
    \centering
    \caption{Generic Meeting Protocols Schema ($M_C$).}
    \begin{tblr}{|c|c|} 
        \toprule
        Features & Description \\
        \midrule
        Meeting ID              & Unique identifier of the meeting \\
        Parliament Number       & The number of parliament of the data \\
        Committee ID            & Unique identifier of the committee \\
        Meeting Title           & The title of the meeting \\
        Date                    & The time of the meeting\\
        Description*            & Description of the meeting (content) \\
        Attendees               & List of parliaments members that attended  the meeting \\
        Number of Attendees     & Number of parliaments members that attended the meeting \\
        Speaker Name            & The ID of the Parliament Member that initiated the meeting \\
        \bottomrule
  \end{tblr}
  \label{tab:protocols_schema}
\end{table*}

\begin{table*}[hbtp]
    \centering
    \caption{The schema for the enriched dataset ($E_C$).}
    \begin{tblr}{|c|c|} 
        \toprule
        Features & Description \\
        \midrule
        Index               &  Index in the dataset \\
        Country             &  Country name  \\
        Vote ID             & Unique identifier of the vote \\ 
        Parliament Number   & The number of the parliament occurred \\ 
        Parliament Type*    & The type of the parliament \\ 
        Session ID          & Unique identifier of the voting session \\
        Vote Date           & The date of the vote \\
        Total For           & Number of members who voted for the proposal\\
        Total Against       & Number of members who voted against the proposal\\
        Member ID           & ID of the Parliament member\\
        Member Full Name    & Parliament member full name\\
        Party Name          &  Name of member's party \\
        Gender              & Gender of the parliament member (Male or Female) \\
        Email               & E-Mail of the parliament member \\
        Party ID            & Unique identifier of the  member's party\\
        Party Name          & Party of the parliament member \\
        Importance Rank     & Number of the importance of the position of parliament member \\
        Is in Alliance      & boolean if the party is in the alliance \\
        Is Current          &  Boolean indicating if the member is in the parliament now\\ 
        Bill ID             & Unique identifier of the person \\
        Title               & The title of the bill \\
        Bill Embedding*            & Embedding of the bill description \\
        Committee ID*       & The ID of the Committee which the bill was proposed \\
        Attendees*               & List of parliaments members that attended  the meeting \\
        Number of Attendees*     & Number of parliaments members that attended the meeting \\
        Speaker Name*            & The ID of the Parliament Member that initiated the meeting \\
        Count of references*        & count of references of the bill’s subject in the parsed protocols data \\
        Vote Result         &  The result of the type (1-for, 2-against, 3-abstain, 4-did not vote) \\
        \bottomrule
  \end{tblr}
  \label{tab:enriched_schema}
\end{table*}

\begin{table*}[hbtp]
    \caption{Experiment results of vote prediction - Accuracy \& F1 Score}
    \resizebox{\textwidth}{!}{
    \begin{tabular}{|c|cc|cc|cc|cc|cc|cc|}
        \toprule
          Country & \multicolumn{2}{c|}{XGBoost} & \multicolumn{2}{c|}{Decision Tree} & \multicolumn{2}{c|}{Random Forest} & \multicolumn{2}{c|}{MLP Classifier} & \multicolumn{2}{c|}{Naïve Bayes} & \multicolumn{2}{c|}{SVM}\\ 
        \midrule
                  & Accuracy & F1 Score & Accuracy & F1 Score & Accuracy & F1 Score & Accuracy & F1 Score & Accuracy & F1 Score & Accuracy & F1 Score \\ 
        \midrule
             Canada & $\textbf{79.807\%}$ & $\textbf{79.658\%}$ & $75.461\%$ & $63.157\%$ & $76.109\%$ & $75.552\%$ & $74.082\%$ & $60.005\%$ & $73.495\%$ & $72.081\%$ & $72.104\%$ & $69.822\%$ \\ 
           Israel &  $\textbf{85.166\%}$ & $\textbf{83.888\%}$ & $73.041\%$ & $72.881\%$ & $79.553\%$ & $78.841\%$  & $69.306\%$ & $62.384\%$ & $67.271\%$ & $66.665\%$ & $65.796\%$ & $61.664\%$  \\ 
          Tunisia  & $\textbf{78.342\%}$ & $\textbf{82.374\%}$ & $73.704\%$ & $65.638\%$ & $75.652\%$ & $76.147\%$ & $53.324\%$ & $42.833\%$ & $53.166\%$ & $47.907\%$ & $53.202\%$ & $48.126\%$   \\
           United Kingdom & $\textbf{80.308\%}$ & $\textbf{80.314\%}$ & $65.010\%$ & $64.504\%$ &  $73.271\%$ & $68.606\%$  & $64.771\%$ & $59.431\%$ & $67.774\%$ & $66.313\%$ & $61.893\%$ & $59.265\%$ \\
           United States &  $\textbf{80.783\%}$ & $\textbf{77.639\%}$ & $75.783\%$ & $73.730\%$ & $79.147\%$ & $78.834\%$ & $75.055\%$  & $71.313\%$ & $72.542\%$ & $70.137\%$ & $67.826\%$ & $64.884\%$  \\
        \bottomrule
    \end{tabular}
    }
    \label{table:results_table}
\end{table*}

\begin{table*}[hbtp]
    \centering
    \caption{Matrix of Overall Bill Prediction with Passed/Rejected Separation.}
     \resizebox{\textwidth}{!} {
    \begin{tblr}{|c|c|c|c|c|c|}
        \toprule
        Country & Total Number of Bills & Successful Passed Bills & Failed Passed Votes & Successful Rejected Bills & Failed Rejected Bills \\
        \midrule
        \textbf{Canada} & 411 & 18 & 66 & 320 & 7 \\
        \textbf{Israel} & 2713 & 911 & 31 & 1347 & 424 \\
        \textbf{Tunisia} & 367 & 137 & 4 & 166 & 60 \\
         \textbf{United Kingdom} & 437 & 188 & 44 & 181 & 24 \\
        \textbf{United States} & 302 & 228 & 22 & 22 & 30\\
        \bottomrule
  \end{tblr}
  }
    \label{table:bill_prediction_results_by_passed}
    \end{table*}
%----------------------------------------------------------------------------------------

\end{document}